\documentstyle[twocolumn,aps,epsfig]{revtex}

\newcommand{\lsim}{\mathrel{\raisebox{-.6ex}{$\stackrel{\textstyle<}{\sim}$}}}

\begin{document}
\draft

\title{
High energy neutrino signals of four neutrino mixing 
}
\author{Sharada Iyer Dutta$^1$, Mary Hall Reno$^{2}$ and Ina Sarcevic$^1$}
\address{
$^1$Department of Physics, University of Arizona, Tucson, Arizona
85721\\
$^2$Department of Physics and Astronomy, University of Iowa, Iowa City,
Iowa 52242
}

\wideabs{
\maketitle
\begin{abstract}
\widetext 
We evaluate the upward shower and muon event rates for two
characteristic four neutrino mixing models for extragalactic
neutrinos, as well as for the atmospheric neutrinos, with energy
thresholds of 1 TeV, 10 TeV and 100 TeV.  We show that by comparing
the shower to muon event rates, one can distinguish between
oscillation and no-oscillation models.  By measuring shower and muon
event rates for energy thresholds of 10 TeV and 100 TeV, and by
considering their ratio, it is possible to use extragalactic neutrino
sources to determine the type of four-flavor mixing pattern.  We find
that over several years of data taking, a kilometer-size detector has
a very good chance of providing valuable information about the physics
beyond the Standard Model.
\end{abstract}}
\vskip 0.1true in

\narrowtext
\section{Introduction}

The combined results of solar, atmospheric and laboratory experiments
with neutrinos, taken at face value, require a fourth neutrino
species. This follows from the observation that the results of the
three categories of experiments require at least three mass-squared
differences $\delta m^2$.  The mass-squared difference for solar
neutrino experiments is limited to $\delta m^2_{solar}\lsim 10^{-3}$
eV$^2$ \cite{Suzuki:2001xw}. The SuperKamiokande results
\cite{Fukuda:1998mi} for atmospheric neutrinos require $\delta
m^2_{atm}\sim 3\times 10^{-3}$ eV$^2$, and the laboratory LSND
experiment \cite{Athanassopoulos:1998pv} limits $\delta
m^2_{LSND}>0.2$ eV$^2$.  While precision measurements of the invisible
decay width of the $Z^0$ boson constrain the addition of a fourth
generation neutrino species with weak interactions, they do not
constrain the possibility of a sterile neutrino species mixing with
the ordinary electron, muon and tau neutrinos.  Analyses of solar and
atmospheric experiments assuming pure $\nu_i\rightarrow \nu_s$
oscillations indicate that the data do not support the hypothesis,
however, combined fits to all of the experimental data do require a
sterile neutrino species \cite{Barger:2000ch}.

Recent SuperK data on atmospheric neutrinos indicate
$\nu_\mu\rightarrow \nu_\tau$ oscillations with mixing being nearly
bi-maximal \cite{Fukuda:1998mi}.  We have shown in Ref.
\cite{Dutta:2000jv} that for extragalactic sources of muon neutrinos,
mixing with tau neutrinos in transit to the Earth leads to distinct
signatures of oscillation which do not require explicit identification
of a tau lepton in large underground experiments. In this paper, we
investigate the signatures of neutrino oscillations in large
underground experiments for models with three active neutrinos and one
sterile neutrino whose mixing is constrained by lower energy
experiments.

\section{Mixing Models and Extragalactic Fluxes}

Global fits to oscillation data fall into two distinct patterns of
neutrino mixing. The first is a mass spectrum in which
$\nu_e\leftrightarrow \nu_s$ with a mass splitting $\delta
m^2_{solar}$, $\nu_\mu\leftrightarrow \nu_\tau$ with $\delta
m^2_{atm}$ and a splitting between the two nearly degenerate pairs
characterized by $\delta m^2_{LSND}$. The approximate mixing matrix in
this scenario has the form \cite{Barger:2000ch}
\begin{equation}
\pmatrix{\nu_s \cr \nu_e\cr \nu_\mu\cr \nu_\tau}= \pmatrix{{1\over
\sqrt{2}} & {1\over \sqrt{2}} & 0 & 0\cr -{1\over \sqrt{2}}& {1\over
\sqrt{2}} & \epsilon & \epsilon\cr \epsilon & -\epsilon & {1\over
\sqrt{2}}& {1\over \sqrt{2}}\cr 0& 0 & -{1\over \sqrt{2}}& {1\over
\sqrt{2}}} \pmatrix{\nu_0 \cr \nu_2\cr \nu_2\cr \nu_3}\ ,
\end{equation}
where $\epsilon < 0.1$ \cite{Barger:2000ch}. Following Ref.
\cite{Barger:2000ch}, we call this a $2+2$ scenario.

The second pattern of neutrino mass and mixing, allowed by the most recent
analysis of the LSND experiment, has large mixing between the three ordinary
neutrinos and a small mixing with the sterile neutrino, characterized by 
small parameters $\epsilon$ and $\delta$. A characteristic mixing matrix
has the form \cite{Barger:2000ch}
\begin{equation}
\pmatrix{\nu_s \cr \nu_e\cr  \nu_\mu\cr \nu_\tau}=
\pmatrix{ 1 & {\delta\over 2}-{\epsilon \over \sqrt{2}} & -{\delta\over 2}
-{\epsilon \over \sqrt{2}} & -{\delta\over \sqrt{2}}\cr
\epsilon & {1\over \sqrt{2}}&  {1\over \sqrt{2}} & 0 \cr
\delta & -{1\over 2}& {1\over 2}& {1\over \sqrt{2}}\cr
0&  {1\over {2}}& -{1\over 2}& {1\over \sqrt{2}}}
\pmatrix{\nu_0 \cr \nu_2\cr  \nu_2\cr \nu_3}\ .
\end{equation}
Following Ref. \cite{Barger:2000ch}, we call this a $1+3$ scenario.

For extragalactic sources, because of the large distances involved,
the oscillation flavor ratios are essentially independent of
mass-squared differences and neutrino energy, since
\begin{equation}
\langle \sin^2\Biggl({1.27 \delta m^2 L\over E}\Biggr)\rangle \simeq
{1\over2}\ ,
\end{equation}
for $\delta m^2$ in eV$^2$, $L$ in km, and $E$ in GeV.  Consequently,
the oscillation probabilities can be written as
\begin{equation}
P(\nu_\ell '\rightarrow\nu_{\ell })= \sum_j\mid U_{\ell j}\mid^2 \mid
U_{\ell ' j}\mid^2\ ,
\end{equation}
in terms of the elements of the neutrino mixing matrix $U_{\ell j}$.
The $\nu_\ell$ flux in detectors $F_\ell^D$, in terms of the fluxes at
the source $F_\ell^S$ are \cite{Ahluwalia:2000fq,Athar:2000yw}
\begin{equation}
F_{\nu_\ell}^D = \sum_{\ell '}P(\nu_\ell '\rightarrow\nu_{\ell })
F_{\nu_{\ell '}}^S\ .
\end{equation}
As we discuss below, our ``source'' fluxes are actually summed over
many sources to yield isotropic fluxes. We continue to denote these
isotropic fluxes, unmodified by oscillations, as the source fluxes.

Given source ratios of fluxes
$\nu_s^S:\nu_e^S:\nu_\mu^S:\nu_\tau^S=0:1:2:0$ yield different ratios
of neutrino fluxes at the detector, depending on whether the $2+2$ or
$1+3$ scenario describes four-neutrino mixing. In the $1+3$ case with
small $\epsilon$ and $\delta$, the detector ratios are approximately
$0:1:1:1$, while for the $2+2$ case, sterile neutrinos make an
important component of the flux at the Earth, with
$\nu_s^D:\nu_e^D:\nu_\mu^D:\nu_\tau^D\simeq 0.5:0.5:1:1$. With three
neutrino species and bi-maximal mixing, one finds $\nu_e^D:\nu_\mu^D:\
\nu_\tau^D\simeq 1:1:1$. Without mixing, the source fluxes and
detector fluxes have the same flavor ratios of $1:2:0$.  As a result,
the $1+3$ scenarios essentially reproduce the three-flavor bi-maximal
mixing scenario, while the $2+2$ scenarios lie between the 3-flavor
bi-maximal mixing model and the no-mixing model.

There are a variety of predictions for isotropic neutrino fluxes from
active galactic nuclei (AGN), gamma ray bursters (GRB) and models with
topological defects.  A sample of these predictions for muon neutrino
plus antineutrino fluxes, in the absence of oscillations, are shown in
Fig. 1. These include the AGN models of Stecker and Salamon (AGN\_SS)
\cite{Stecker:1996th} and of Mannheim Model A (AGN\_M95)
\cite{Mannheim:1995mm}. 
Both of these models predict neutrinos fluxes that
represent the upper bounds for their class of the
models.
In particular, the Stecker-Salamon flux is an upper bound for AGN core
emission, while Mannheim Model A is an upper bound for AGN jet
emission models.  The Stecker-Salamon flux is bound by the the diffuse
X-ray background, while Mannheim flux is bound by the extragalactic
gamma ray background.  GRB predictions are represented by the
Waxman-Bahcall model of Ref. \cite{Waxman:1999yy} (GRB\_WB).  Two
topological defect models, from Sigl, Lee, Schramm and Coppi
(TD\_SLSC) \cite{Sigl:1997gm} and Wichoski, MacGibbon and
Brandenberger (TD\_WMB) \cite{Wichoski:1998kh} are also shown.  We
include fluxes with energy behaviors like $1/E$ and $1/E^2$ as well.
We chose normalization for each of these fluxes consistent with the
current limits \cite{Andres:2000jf}, $F_{\nu_\mu}^S=10^{-12}\ (E/{\rm
GeV})^{-1}$GeV$^{-1}$cm$^{-2}$sr$^{-1}$s$ ^{-1}$ and
$F_{\nu_\mu}^S=10^{-6}\ (E/{\rm
GeV})^{-2}$GeV$^{-1}$cm$^{-2}$sr$^{-1}$s$^{-1}$ (in which the fluxes
include neutrinos and antineutrinos in equal amounts and correspond to
the initial fluxes before accounting for oscillations).  Our choice of
normalization for $E^{-2}$ flux is a factor of ten larger than
recently proposed neutrino flux associated with GRB (
$F_{\nu_\mu}^S=10^{-7}\ (E/{\rm
GeV})^{-2}$GeV$^{-1}$cm$^{-2}$sr$^{-1}$s$^{-1}$) \cite{Waxman}, and a
factor of 50 larger than the upper bound for strong source evolution
previously discussed by Waxman and Bahcall \cite{bounds}.  The
$E^{-1}$ flux is smoothly cut off at high energies, as indicated in
Fig. 1.

Although the normalizations of the fluxes of neutrinos from astrophysical
sources are uncertain, we evaluate event rates quantitatively 
for these representative energy behaviors of the fluxes which, 
as models
improve, can be rescaled to accommodate different normalizations.

\begin{figure}[!hbt]
\rule{0.0cm}{0.0cm}\vspace{-0.5cm}\\\\
\epsfxsize=7.5cm
\epsfbox[0 0  4096 4096]{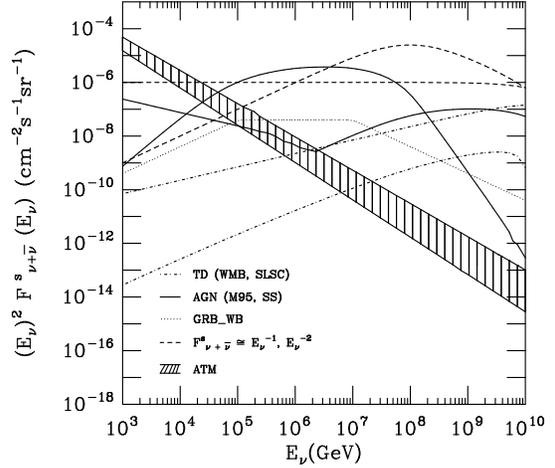}
\caption{Isotropic muon neutrino plus antineutrino flux predictions
for AGN models (solid lines, upper curve at low energy corresponds to
AGN\_M95, while the lower curve is for AGN\_SS model), GRB (dotted
line), topological defects models (dash-dotted lines, upper curve
corresponds to TD\_(Model A) WMB, while the lower curve is for
TD\_SLSC), $E^{-1}$ flux (lower dashed line at low energy) and
$E^{-2}$ (upper dashed line at low energy) and angle-dependent
atmospheric (ATM). The flux is scaled by neutrino energy squared and
the antineutrino flux is taken equal to the neutrino flux.}
\end{figure}

For comparison, the hatched curve shows the angle dependent
atmospheric flux (ATM) \cite{Agrawal:1996gk} from kaon and pion decays
(conventional flux). We have extrapolated this flux beyond the 10 TeV
range given in Ref. \cite{Agrawal:1996gk} using an angular dependent
power law.  This does not account for the change in the input cosmic
ray spectrum, the so-called ``knee'' at energies of $\sim 10^6$ GeV
\cite{Gaisserbook}, which translates to a reduced conventional
neutrino flux at high energies.  Thus, we overestimate the
conventional neutrino event rates, especially at our highest threshold
of 100 TeV, however, at that threshold, the atmospheric rates are
small.  Our evaluation does not include the prompt neutrino flux
from semileptonic decays of charmed particles produced in the
atmosphere, which result in neutrino fluxes with a power law increased
by one factor of energy. Recent evaluations of the prompt neutrino
flux suggest that it is important only above 100 TeV
\cite{Pasquali:1999ji,Gelmini:2000ve,Gondolo:1996fq}, although it has
been emphasized that there are large theoretical uncertainties in the
evaluation \cite{Costa:2001fb}.

In our previous work \cite{Dutta:2000jv}, we have evaluated how
neutrino interactions in the Earth modify these representative
neutrino fluxes with the regeneration attributed to the neutral
current interactions.  In addition, for tau neutrinos, we described
the extent to which neutral current interactions and charged current
production of $\tau$ followed by its decay, regenerate
neutrinos. There, we made a detailed numerical evaluation of the
``pile-up'' of tau neutrinos that was discussed by Halzen and
Saltzberg in Ref. \cite{Halzen:1998be}.  In the next section, we use
the neutrino fluxes of Fig. 1 modified due to their passage through
the Earth \cite{Dutta:2000jv} in the appropriate flavor proportions
for the $1+3$ and $2+2$ scenarios.

\section{Underground Signatures}

Backgrounds from atmospheric muons make downward event rates difficult
to extract, so our focus is on upward events. Two types of events will
be produced: muon events and shower events.  The muon events come from
upward muons from $\nu_\mu\rightarrow \mu$ charged current events and
from charged current production of taus followed by a muonic decay,
$\nu_\tau\rightarrow\tau\rightarrow \mu X$.  In spite of the pile-up
in the tau neutrino flux, the net effect of oscillations is to reduce
the muonic event rate by approximately a factor of two relative to the
no-oscillation rate, whether in the $1+3$ or $2+2$ mixing
scenario. Given the uncertainties in the normalizations of the
extragalactic fluxes (unlike the atmospheric flux), the muon event
rate is not enough to distinguish oscillation scenarios from
no-oscillation scenarios.

The distinction between oscillation and no-oscillation scenarios comes
by comparing the muon rate with the shower rate. The principle of this
procedure is very similar to what some experiments will be able to do
with low energy neutrinos: namely, to compare the neutral current rate
to the charge current interaction rate. In large neutrino telescopes,
tau identification is difficult, and we assume that 
showers of hadronic and electromagnetic origin cannot be distinguished.
Consequently, the shower rate comes from the following processes
from neutrino-nucleon ($N$) interactions:
\begin{equation}
\matrix{
\nu_\tau N\rightarrow \tau+{\rm hadrons},\ \tau\rightarrow \nu_\tau
+{\rm hadrons}\ ,\cr
\nu_\tau N\rightarrow \tau+{\rm hadrons},\ \tau\rightarrow \nu_\tau
+e+\nu_e\ ,\cr
\nu_\tau N\rightarrow \nu_\tau+{\rm hadrons},\cr
\nu_{e,\mu} N\rightarrow \nu_{e,\mu}+{\rm hadrons},\cr
\nu_{e} N\rightarrow {e}+{\rm hadrons}\ .}
\end{equation}
The shower energy is taken as the sum of hadronic and electron energies
including both the production vertex and the decay vertex, where applicable.

The upward muon rates are determined from an evaluation of
the quantity
\begin{eqnarray}
{\rm Rate} &&= A N_A\int_{E_\mu^{\rm min}} dE_\nu
\int dy \int dz \langle R_\mu(E_\mu, E_\mu^{\rm min} )\rangle
{d\sigma_{cc}\over d y}
\\ \nonumber
&& \times F_\nu(E_\nu, X)\Theta (E_\mu-E_\mu^{\rm min})\, f_\mu (E_\nu,y,z)
\end{eqnarray}
where $E_\mu=E_\nu(1-y)z$.
The function $f$ depends whether the source of the muons is 
muon neutrino charged current interactions or tau neutrino charged current
interactions followed by the muonic decay of the tau:
\begin{eqnarray}
f_\mu(E_\nu,y,z)&&=\delta(1-z)\ {\rm for \ }\nu_\mu\rightarrow \mu\ ,\\ \nonumber
f_\mu(E_\nu,y,z)&&={dn(E_\mu)\over dz}\ {\rm for \ }\nu_\tau\rightarrow \tau
\rightarrow \mu\ ,
\end{eqnarray}
where the decay formula appears in the appendix of
Ref. \cite{Dutta:2000jv}.  The quantity $A$ is the effective area
(taken as 1 km$^2$ here), $N_A$ is Avogadro's number and the
differential cross section is for neutrino-
interactions.  The average range of the muon $\langle R\rangle$
\cite{Lipari:1991ut}, for initial energy $E_\mu$ and final energy
$E_\mu^{\rm min}$ is the additional space dimension that makes up the
target volume. The column depth ($X$) dependence of the neutrino flux
represents the angular dependence of the flux due to the attenuation
of the flux after passage through the Earth, and in the case of
atmospheric neutrinos, includes the angular dependence of the flux at
the surface of the Earth.

For the shower rate, one does not have the benefit of the muon range,
so the full instrumented volume $V$ enters into the calculation of the event
rate, here taken to be 1 km$^3$:
\begin{eqnarray}
{\rm Rate} &&= V N_A\int_{E_{\rm shr}^{\rm min}} dE_\nu
\int dy \int dz 
{d\sigma\over d y}
\\ \nonumber
&& \times F_\nu(E_\nu, X)\Theta (E_{\rm shr}-E_{\rm shr}^{\rm min})\, 
f_s(E_\nu,y,z)\ ,
\end{eqnarray}
where $f_s$ has a similar form to $f_\mu$. Details of the differential
decay distribution of the tau used in the evaluation, as well as the 
specific dependence of $E_{\rm shr}$ on $E_\nu, y$ and $z$ appear
in \cite{Dutta:2000jv}. The differential cross section is either charged
current or neutral current, depending on which process in
Eq. (6) is being considered.

The neutrino and antineutrino cross sections \cite{Gandhi:1998ri} are
evaluated using the CTEQ5 parton distribution functions
\cite{Lai:2000wy}, and the attenuation of the fluxes assume Earth
densities of the Preliminary Earth model described in
Ref. \cite{prem}.  As in Ref. \cite{Dutta:2000jv}, our upper bound of
integration is $10^6$ GeV because the attenuation of $\nu_\tau$ fluxes
at higher energies have not been evaluated, due to the complication of
including tau energy loss at high energies \cite{Dutta:2001hh}. Our
results for the 100 TeV threshold are therefore conservative.

The ratio of the shower rate to the muon rate for the extragalactic
flux models of Fig. 1 are used to bracket the theoretical expectations
for the ratio in $2+2$ and $1+3$ mixing models, and compared with the
standard model expectations. As in Ref. \cite{Dutta:2000jv}, we
separate the flux models into two categories: more steeply falling
fluxes ($E^{-2}$, ATM and the Mannheim AGN flux) and the less steep
behavior of the other five sample fluxes of Fig. 1. Experimentally,
the steeply falling fluxes and the models with a weaker energy 
dependence are characterized by distinct muon event rate
nadir angle dependence and the threshold energy dependence.  
The two separate categories of fluxes, for threshold energies of 1, 10 and 100
TeV are shown in Figs. 2 (a-f).  The $1+3$ models are bracketed by the
black lines, the $2+2$ models are represented by the shaded area and the 
standard model with the dark band.

For the 1 TeV threshold, the $1+3$ and $2+2$ event rates as a function
of nadir angle overlap for both energy behaviors of the input fluxes.
The no-oscillation band lies below the oscillation bands with the
exception of the near horizontal rates in the steeply falling flux
category (labeled $E^{-2}$). The separation between the oscillation
and no-oscillation scenarios is better for higher thresholds, looking
only in terms of the theoretical ratio of shower to muon event
rates. However, the higher thresholds have lower event rates overall.

The bands in Fig. 2 indicate the different ratios for oscillation
models relative to no-oscillation models, however, statistical errors
from the signal as well as from the atmospheric background need to be
included in order to interpret whether or not oscillation scenarios
are distinguishable.  In the next section, we include a discussion of
statistical errors.

\begin{figure}[!hbt]
\rule{0.0cm}{0.0cm}\vspace{3cm}\\
\epsfxsize=7.5cm
\epsfbox[0 0  4096 4096]{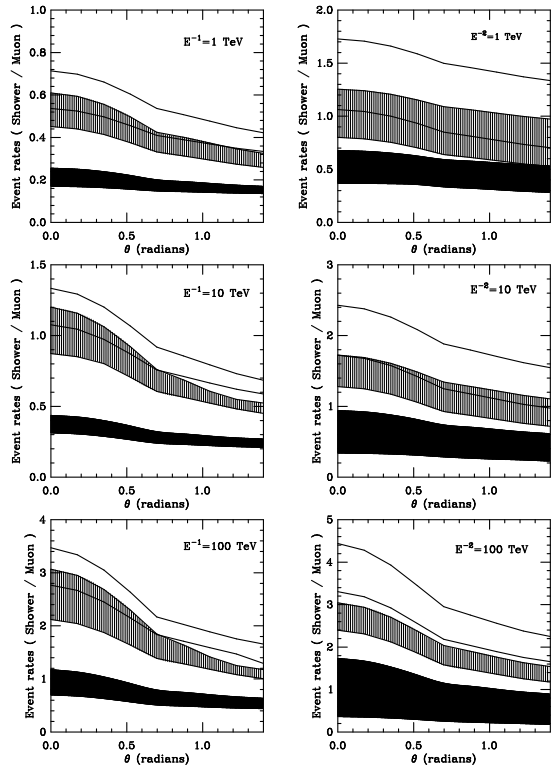}
\vspace{.5cm}\\
\caption{Ratio of shower event rates to muon event rates for the
$1+3$, $2+2$ and no oscillation scenarios as a function of nadir
angles for threshold energies of 1 TeV, 10 TeV and 100 TeV, for the
indicated fluxes. The black lines bracket the $1+3$ scenario, the 
shaded area represents the $2+2$ scenario and the dark band
represents no-mixing case for the given fluxes.}
\end{figure}

\section{Discussion}

In Tables I-III we present our results for the shower and muon event
rates integrated over the nadir angle for energy thresholds of 1 TeV,
10 TeV and 100 TeV and for a kilometer-size detector.  
The no-oscillation results agree with evaluations in Ref. \cite{Gandhi:1998ri}
for fluxes in common, modulo corrections due to different parton
distribution functions, a more precise evaluation of neutrino attenuation
in the Earth \cite{Naumov:1999sf} and the upper limit of energy integration. 
We do not include the topological defect models in these tables because the
event rates are so low. For the 1 TeV threshold, only $0.1-0.2$ upward
muon events are predicted per year for a km$^2$ instrumented area for
the TD\_WMB model. In all cases, we have considered shower rates in a
km$^3$ volume, and muon rates/km$^2$.
The two oscillation scenarios give the same muon event rates but very
different shower rates.  From Table I, we note that for 1 TeV energy
threshold, atmospheric background is large.  

\begin{figure}[!hbt]
\rule{0.0cm}{0.0cm}\\
\epsfxsize=7.5cm
\epsfbox[0 0  4096 4096]{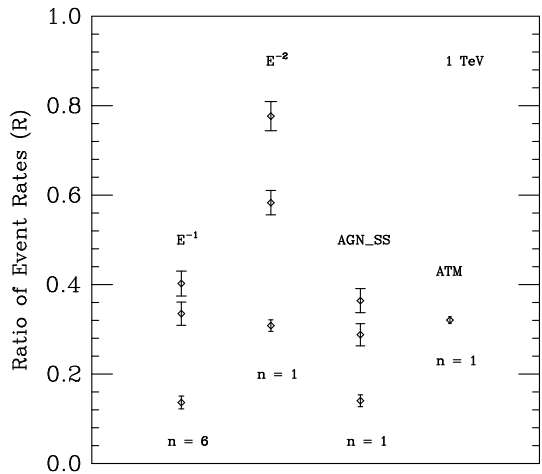}
\caption{Integrated over solid angle, the ratio of upward
shower to muon event rates for given flux models for a threshold energy
of 1 TeV. The error bars for $1+3$ scenario (top data), $2+2$ scenario
(middle data), and no-mixing case (bottom data) are determined
assuming $n$ years of data taking.}
\end{figure}

In Figs. 3-5, we show the ratio $R$ of the shower to muon event rates
integrated over nadir angles ($R\equiv N_{shr}/N_\mu$) for energy
thresholds of 1 TeV, 10 TeV and 100 TeV for a variety of fluxes and
for different oscillation scenarios.  The ratio of event rates in
Figs. 3-5 have the atmospheric flux subtracted.  The errors in the
shower and muon rates ($\sigma_{shr}$ and $\sigma_\mu$) are evaluated
with the assumption of a Poisson distribution, including the
statistical error on the number of background events from the
atmospheric flux. The error in the ratio of event rates is then given
by,
\begin{equation}
\sigma_R=R\sqrt{{\sigma_{shr}\over N_{shr}}+{\sigma_\mu\over N_\mu}}\ .
\end{equation}
The event rates $N_{shr}$ and $N_\mu$ are from Tables I-III, 
scaled by $n$ years as indicated in Figs. 3-5. 
 
From Fig. 3 and Table I, we note that AGN fluxes, as well as $E^{-2}$
and $E^{-1}$ fluxes, would have the best possibility of separating
different oscillation scenarios by detecting the showers and muons in
kilometer-size detector over the period of one to six years, assuming
a good understanding of the atmospheric muon neutrino and electron
neutrino fluxes and normalizations near their current upper bounds.
After one to six years of data taking, one would be able to separate
oscillation from no oscillation scenario for the fluxes in Fig. 3.
The GRB\_WB flux would require more than twenty years of data to
reduce statistical errors so that the error bars do not overlap for
the two oscillation scenarios.

\begin{figure}[!hbt]
\rule{0.0cm}{0.0cm}\\
\epsfxsize=7.5cm
\epsfbox[0 0  4096 4096]{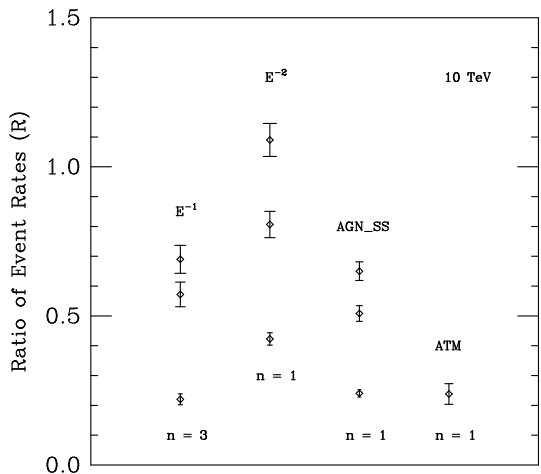}
\caption{Integrated over solid angle, the ratio of upward
shower to muon event rates for given flux models for a threshold energy
of 10 TeV.The error bars for $1+3$ scenario (top data), $2+2$ scenario (middle data),
and no-mixing case (bottom data) are determined assuming $n$ years of
data taking.}
\end{figure}

The ratio of shower to muon event rates for an energy threshold of 10
TeV are shown in Fig. 4.  As seen in Table II, the AGN model of
Stecker and Salamon predicts large shower and muon rates, an order of
magnitude larger than the atmospheric background.  The AGN model of
Mannheim and the GRB model of Waxman and Bahcall predict rates to be
about 15-40$\%$ of the background, while $E^{-2}$ and $E^{-1}$ fluxes
give significantly more events than the atmospheric background when
oscillations are taken into account, assuming that the fluxes occur
with a normalization at their current upper limit.  Thus, there is a
possibility of detecting extragalactic neutrinos by imposing a 10 TeV
energy threshold [5].  We note that for the AGN\_SS flux, as well as
for $E^{-1}$, $E^{-2}$ fluxes, one to three years would be sufficient
to separate different oscillation scenarios. For the AGN\_M95 and
GRB\_WB fluxes, it would be necessary to take data for seven to nine
years respectively.  The 10 TeV threshold is generally more favorable
than the 1 TeV threshold because of the reduction of the atmospheric
background and a sufficiently large signal.

If the energy threshold is increased to 100 TeV, the atmospheric
background is small.  The shower and muon event rates are several
hundreds for AGN\_SS model, while for all other models the rates are
much smaller, but still significant, for kilometer-size detector in a
time period of one year.  Furthermore, the ratio of the shower and
muon events can clearly separate oscillation and no-oscillation
scenarios, as well as distinguish two oscillations scenarios that we
consider.  In Fig. 5 we show this ratio including statistical errors.
We find that even with error bars it still seems possible to
distinguish between different oscillation scenarios.  However, in
order to reduce statistical errors, the GRB\_WB and AGN\_M95 flux
predictions would require collecting data for more than a decade.  On
the other hand, one year is sufficient for the AGN\_SS flux.  In the
case of an $E^{-1}$ flux, we expect several hundreds of showers and
muons over the period of three years, which is necessary in order to
reduce error bars to the point of being able to distinguish
oscillation scenarios.  For a steeper flux, $E^{-2}$, one year is
sufficient to get enough events as well as to reduce the statistical
errors at the current AMANDA limit on the isotropic flux
normalization.

\begin{figure}[!hbt]
\rule{0.0cm}{0.0cm}\\
\epsfxsize=7.5cm
\epsfbox[0 0  4096 4096]{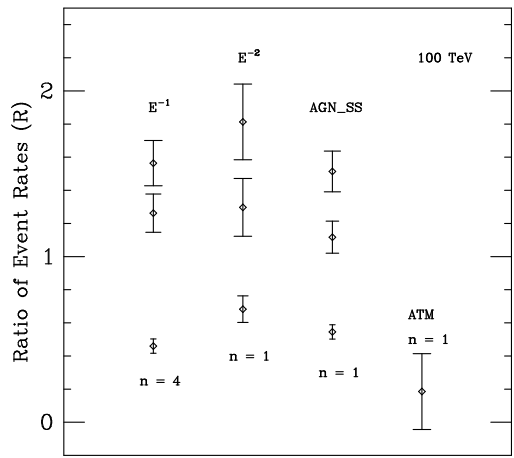}
\caption{Integrated over solid angle, the ratio of upward
shower to muon event rates for given flux models for a threshold energy
of 100 TeV.  The error bars for $1+3$ scenario (top data), $2+2$ scenario (middle data),
and no-mixing case (bottom data) are determined assuming $n$ years of
data taking.}
\end{figure}

Figs. 3-5 show the feature illustrated by Fig. 2, when specific fluxes are
integrated over nadir angle. Fluxes with similar spectra cluster with the
same ratio, here, the Stecker-Salamon flux and $E^{-1}$ have similar
energy behaviors and similar ratios. Therefore, by using muon event
energy and nadir angle dependence to characterize the energy spectrum of
the incident flux, even without a detailed knowledge of the source of
neutrinos, the ratio $R$ will point to whether or not muon neutrinos mix
with tau neutrinos in one of the two schemes described here.

The ratios in Figs. 3-5 differ for the different
energy dependences of the  fluxes for several reasons.  The
muon range increases with energy, so the muon event rates are
sensitive to the energy dependence in a different way than the shower
rates which have a constant volume factor.  For less steep fluxes,
{\it e.g.}, $E^{-1}$ and AGN\_SS, the high energy part of the spectrum
gives a larger contribution to the muon rate than for fluxes with a
steeper energy behavior, {\it e.g.}, $E^{-2}$.  The ratio $R$ of
shower to muon event rates tends to increase with the steeper fluxes,
and tends to increase with energy threshold.  For all except the
atmospheric flux, the electron neutrino contribution to the shower
rate is substantial. The electron neutrino component of the
atmospheric flux is small because of the high energies considered
here, so the atmospheric ratio is not directly comparable to the
ratios for other fluxes.

\section{Conclusion}

We have shown that measurements of the upward showers and muons, and
considering their ratio, with the kilometer-size detector, could
provide very good test of different oscillation scenarios. Taking into 
account statistical errors, we find that generic fluxes, such as 
$F_{\nu_\mu}^S=10^{-12}\ (E/{\rm GeV})^{-1}$GeV$^{-1}$cm$^{-2}$sr$^{-1}$s$^{-1}$ and
$F_{\nu_\mu}^S=10^{-6}\ (E/{\rm GeV})^{-2}$GeV$^{-1}$cm$^{-2}$sr$^{-1}
$s$^{-1}$, with energy threshold of 10 TeV and 100 TeV, would have
enough statistics in one to four years of data taking to distinguish
not only between oscillation and no oscillation scenario, but also
whether it is 2+2 or 1+3 model.  The AGN model of Stecker-Salamon
\cite{Stecker:1996th}, with a 10 TeV energy threshold predicts between
1200 and 2000 muons/km$^2$ per year and between 500 and 800 shower
events/km$^3$ per year, with clear separation between different
oscillation scenarios.  The other AGN model, proposed by Mannheim
\cite{Mannheim:1995mm}, predicts less events and would require seven 
years to reduce statistical errors in the ratio of shower to muon
event rates such that there is clear distinction between not only the
no oscillation and oscillation cases, but also different oscillation
models.  If we want to use GRB\_WB flux \cite{Waxman:1999yy} to test
these models, it would take nine years with the kilometer-size
detector to get sufficient statistics.  A higher energy threshold of
100 TeV makes the atmospheric neutrino background small, but
event rates for extragalactic neutrinos are also reduced.  Still, the
AGN\_SS model would be able to provide valuable information about the
oscillation scenario, as well as $E^{-2}$ flux in just one year of
data, while an $E^{-1}$ flux with our normalization would need four 
years.  Other fluxes, such as GRB\_WB and AGN\_M95 flux would require
much longer time.

The assumption of the existence of sterile neutrino is necessary to
explain the solar, atmospheric, reactor and accelerator data. The
MiniBooNE experiment {\cite{mini}} will search for $\nu_\mu
\rightarrow \nu_e$ oscillations and test the LSND results.  The
presence of the sterile neutrino in the 2+2 model can be tested by the
measurement of the suppression of NC/CC ratio of the solar neutrinos
by the SNO experiment \cite{SNO} whereas 1+3 model can be tested by
searches from small amplitude oscillations at short baseline
experiment such as ORLaND \cite{orland}.  Since the 1+3 scenario at
high neutrino energy mimics the three flavor bi-maximal mixing between
$\nu_\mu\leftrightarrow \nu_\tau$, the results here can also be
applied to three-flavor models.

Observation of the flavor ratio of extragalactic neutrinos could serve
as a complementary test for oscillation models that incorporate
sterile neutrinos.  Oscillations of neutrinos from astronomical
sources are averaged over very long baseline and thus cannot provide
any direct information about the neutrino mass.  However, we have shown that
extragalactic neutrinos could be used to distinguish between different
oscillation scenarios.  The observed muon and shower event
distributions as a function of angle and energy threshold will help
determine the energy behavior of the incident extragalactic neutrino
flux and point to a model or class of models for their sources.
Combined measurements of the upward showers and muons and, in
particular, their ratio provides a basic test of the oscillation
scenario.

\vskip 0.1true in

\leftline{\bf Acknowledgments}  

\vskip 0.1true in

The work of S.I.D. and I.S. has been supported in part
by the DOE under Contracts 
DE-FG02-95ER40906 and DE-FG03-93ER40792.  
The work of M.H.R. has been supported in
part by
National Science Foundation Grant No.
PHY-9802403.

\vskip 0.1true in

\leftline{\bf Note added:}

\vskip 0.1true in

After the submission of our paper, the SNO collaboration 
has reported their first results \cite{SNO1} which together with
the SuperKamiokande results \cite{Suzuki:2001xw}
favor oscillation into active flavors, however, given the theoretical
and experimental uncertainties, oscillations into 
sterile neutrinos are still a possibility.

\newpage

\onecolumn
\begin{table}
\caption{Integrated upward shower (muon)
rates per year for energy threshold of 1 TeV.}
\begin{tabular}{lccccccc}
Model & $E^{-1}$ & $E^{-2}$ & AGN\_SS & AGN\_M95 & GRB\_WB & ATM & \\
1+3 &  170(422)
      &  2528(3255)
      &  918(2521)
      &  224(157)
      &  26.9(65.8)\\
2+2 &  141(422)
      &  1898(3255)
      &  726(2521)
      &  163(157)
      &  20.9(65.8)\\
No osc &  88.1(646)
      &   1750(5676)
      &  593(4228)
      &  158(280)
      &  18.3(113)
      &  2355(7346)\\
\end{tabular}
\end{table}
\begin{table}
\caption{Integrated upward shower (muon) rates per year for
energy threshold of 10 TeV.}
\begin{tabular}{lccccccc}
Model & $E^{-1}$ & $E^{-2}$ & AGN\_SS & AGN\_M95 & GRB\_WB & ATM & \\
1+3 & 150(217)
      &  954(875)
      &  809(1245)
      &  34.4(20.2)
      & 21.3(28.4)\\
2+2 &  124(217)
      & 706(875)
      & 633(1245)
      & 24.5(20.2)
 & 16.3(28.4)\\
No osc & 75.4(342)
      &  652(1540)
      &  514(2139)
      &  24.4(36.3)
      & 14.2(49.7)
      & 58.5(246)\\

\end{tabular}
\end{table}

\begin{table}
\caption{Integrated upward shower (muon) rates per year for
energy threshold of 100 TeV.}
\begin{tabular}{lccccccc}
Model & $E^{-1}$ & $E^{-2}$ & AGN\_SS & AGN\_M95 & GRB\_WB & ATM & \\
1+3 &   84.8(54.2)
      &   182(100)
      &   386(255)
      &   2.46(0.999)
      &   7.28(4.03)\\
2+2 &  68.4(54.2)
      &  130(100)
      &  285(255)
      &  1.69(0.999)
      &  5.21(4.03)\\
No osc &  42.0(91.3)
      &   124(182)
      &   251(460)
      &   1.78(1.83)
      &   4.99(7.31)
      &  0.774(4.18) \\
\end{tabular}
\end{table}
\end{document}